# "Dark" vs "Bright" Excitons in Carbon Nanotubes: Applications to Quantum Optics

**Nikolai G. Kalugin[1,*] and Yuri V. Rostovtsev[2]**

[1]Materials Engineering Department, New Mexico Tech, 801 Leroy Pl., Socorro NM 87801, USA

[2]Physics Department and Institute for Quantum Studies, Texas A&M University, College Station TX 77843, USA

[*]Corresponding author: nkalugin@nmt.edu

(Dated: December 28, 2008)

Abstract

It is strong Coulomb effects in carbon nanotubes that lead to formation of the so-called "bright" and "dark" (forbidden one-photon optical transition) exciton states, and dramatically decrease the efficiency of one-photon light emission via trapping of the carriers by "dark" states. We demonstrate that the proper use of these "bright" and "dark" exciton states with distinctively different recombination times may turn the situation around: the use of quantum coherence and multiphoton schemes of excitation potentially not only allow one to efficiently manipulate the dark states , but can also create conditions for efficient light generation in different frequency regions, produce "slow" or "fast" light, implement quantum light storage, media with a negative refractive index, and other quantum-optical regimes. Possible quantum-optical carbon nanotube devices have a potential for suitable performance at elevated temperatures, because the binding energies of excitons in single-walled nanotubes are hundreds-of-meV high.

OCIS Codes: 270.0270 Quantum optics, 160.4236 Nanomaterials, 020.1670 Coherent optical effects

In this letter, we have shown that the specific properties of excitons in single-walled carbon nanotubes (SWNT) [1] allow us to apply the well-developed "arsenal" of quantum optics [2,3] for more detailed studies of these excitons. We have proposed several schemes of quantum coherence generation in the exciton system of SWNTs and discussed the conditions of possible "proof-of-principle" experiments. The key idea of our proposal is the application of correlated multiphoton excitation to SWNTs for the achievement of the coherent population trapping related regimes. We have demonstrated that the features of exciton states in semiconductor SWNTs provide the possibility of the implementation of carbon nanotubes as a new material for quantum optics.

Carbon nanotubes show various properties resulting from their one-dimensional (1D) structure [1]. The strong exciton-binding effects in one-dimensional insulators dominate the density of states effects in optical spectra [4-9]. The doubly degenerate nature of the valence and conduction bands in SWNT results in four nearly degenerate excitonic substates for each quantum state of the exciton. Such nearly degenerate excitons are composed of one bright (one-photon allowed) and three dark (one-photon forbidden) excitons [5-9] (see Fig.1).

The radiative recombination of the dark states, which are the lowest energy exciton states in carbon nanotubes, is dramatically suppressed due to the trapping of carriers by "dark" states. The role of "dark" excitons was recently revealed, and several methods of modification SWNT optical properties via the "brightening" of "dark" excitons have been demonstrated [7,8]. Summarizing the existing knowledge about properties of "bright" and "dark" excitons without rigorous detailization of influence of diameters and chiralities on these parameters, we have to point out the known ranges of important parameters: according to time-resolved measurements, the radiative recombination times for "bright" excitons are in hundreds of fs - tens of ps range, the radiative recombination times for "dark" excitons are somewhere between sub-ns - hundreds of ns/microsecond range [6,10]. The nonradiative recombination is dominant in nanotube bundles but may be efficiently suppressed in separated nanotubes.

The "dark"/"bright" splitting is of the order of several meV-tens of meV depends on diameter and chirality [6,8]. The values of dipole moment for optical transitions are of the order of several eA [11], and change depends on diameter and chirality of nanotubes [6]. The binding energies of excitons in SWNTS are of the order of several hundreds of meV [6].The detailed knowledge about exciton states is far from being complete, but the existing data already allows us to make the clear conclusion: carbon nanotubes are a promising medium for quantum optics. Large dipole moments of optical transitions and the existence of "bright" and "dark" states with distinctively different recombination times allow us to propose several multiphoton schemes for quantum coherence-based manipulation of exciton states.

Let us note that the difference between relaxation rates is very unique and promising for applications to quantum optics where the coherence plays a very important role. Quantum coherence effects have been a focus of research activities for the last two decades because they may drastically change the optical properties of a medium. For example, electromagnetically induced transparency (EIT) [2,3,12], predicted and observed in CW and pulsed regimes [13,14], practically allows absorption to vanish. EIT has been successfully demonstrated under different experimental conditions in different atomic, molecular, and solid-state systems at different wavelength regions.

The nonlinear response of a resonant atomic medium at moderate optical intensities can be strongly enhanced by induced quantum coherence between long-lived sublevels of the ground state [13,14]. In order to develop bright sources for efficient generation of IR and FIR/THz pulses [15], it is possible to manipulate a coherent medium and produce optical pulses at rates faster than the relaxation rates of the medium [16]. Also, the nonlinear properties of such media are enhanced, allowing the implementation of quantum light storage [17], nonlinear optics at a few photon level [18,19], and other nonlinear effects [20].

The typical schemes of excitation which can demonstrate coherent effects are shown in Fig.1. The quantum coherence, which is responsible for enhancement of nonlinear effects, is given [3] by

$$\rho_{ab} = i \frac{n_{ab} + \frac{|\Omega|^2}{\Gamma_{cb}\Gamma_{ca}} n_{ca}}{\Gamma_{ab} + \frac{|\Omega|^2}{\Gamma_{cb}}} \Omega_p \tag{1}$$

where $n_a$, $n_b$, and $n_c$ are the populations in levels a, b, and c correspondingly; $n_{ab} = n_a - n_b$, $n_{ca} = n_c - n_a$; $\Omega = dE/\hbar$ is the Rabi frequency for driving transition with the dipole moment $d$; $\Gamma_{ab} = \gamma_{ab} + i\delta\omega_p$; $\Gamma_{ca} = \gamma_{ca} + i\delta\omega_d$; $\Gamma_{cb} = \gamma_{cb} + i(\delta\omega_p - \delta\omega_d)$; $\delta\omega_p = \omega_{ab} - \nu_p$ is the detuning of the probe field; $\delta\omega_d = \omega_{ac} - \nu_d$ is the detuning of the drive field; $\gamma_{ab}$ and $\gamma_{ca}$ are the relaxation rates (inverse relaxation times) for high frequency transitions, and $\gamma_{cb}$ is the relaxation rate of coherence between levels b and c.

As seen from Equation (1), the physics of coherent effects is based on the formation of the so-called coherent population trapping (CPT) state $|\Psi_{CPT}\rangle$, given by

$$|\Psi_{CPT}\rangle = (\Omega_1|b\rangle - \Omega_2|c\rangle)/(\Omega_1^2 + \Omega_2^2)^{1/2} \tag{2}$$

where the Rabi frequencies have been defined above. The excitation of the dark state is limited by the relaxation rates in the system. In order to overcome the relaxation rates, the coherent fields (optical, IR and far-IR) should be strong enough. Namely, the Rabi frequencies should meet the following condition (see details in [3] )

$$\Omega_1^2 + \Omega_2^2 > \gamma_{cb}\Gamma_{ab} \tag{3}$$

where $\gamma_{cb}$ is the relaxation of the CPT state and $\Gamma_{ab}$ is the relaxation of the "optical" transition. For example, for the scheme shown in Fig.1(A), $\Omega_{ab} = \Omega_1$, $\Omega_{ac} = \Omega_2$, $\delta\omega_p = \delta\omega_d = \Delta_1$. The Rabi frequency is related to the radiation intensity $P$ as:

$$\Omega_{\alpha\beta} = d_{\alpha\beta}E/\hbar \approx 5 \cdot 10^9 \, P^{1/2} \tag{4}$$

We can estimate the power that is necessary to achieve the CPT regime in carbon nanotubes (CNTs) taking the relaxation of the long-lived exciton state $\gamma_{cb}^{-1}$ in the range of 1 μs – 30 ns range, and the relaxation at optical transition $\Gamma_{ab}$ of the order of $10^{12}$ s$^{-1}$ [6,10]. For a magnitude of the dipole moment, we use $d_{\alpha\beta}$ ~6 eA [11]. Thus, the power $P$ can be estimated to be larger than a threshold value laying in the range given by the range of relaxation rates known at this time (more research on relaxation times should be done) from 0.2 to 20 W/cm$^2$. For pulsed lasers, this level of power is easily reachable (for example, for a 1 μJ pulse of 1 ps duration, the intensity is $10^6$ W/cm$^2$).

The above estimates show a promise that the coherent effects can be implemented in CNTs. Then, the demonstration of the CPT can be served as a key experiment providing the proof-of-principle demonstration of coherent effects for such objects. The first experiment we suggest is to demonstrate the effect of CPT by observing modulation of fluorescence or optical transmission (using, for example, the scheme (A) of Fig.1). Let us note here that the ratio of $\gamma_{cb}/\Gamma_{ab}$ in carbon nanotubes is even better than in atomic gases [2,3,12]. The use of femtosecond laser pulses allows one to implement stimulated rapid adiabatic passages (STIRAP) [12] in CNTs providing the quantum control of the system [18].

Let us note that the energy threshold required to observe coherent effects can be even less, taking into account that the size of CNTs is very small, meaning the actual energy interacting with CNTs is even smaller. In principle, CNTs can efficiently interact with single photons. To enhance efficiency of interaction between CNTs and radiation, an individual CNT can be placed in an optical resonator (see Fig.2) where, in principle, even interaction with single photons may be sufficient for the achievement of conditions for the observation of quantum-optical effects. This decreases the necessary intensity even further and opens even more applications of CNTs as sources and detectors for single photons.

The demonstration of high efficiency for *quantum coherence-enhanced* nonlinear interaction should be the first step of research (the resonant enhancement of optical nonlinearity in CNTs was recently reported in [24]). As the next step, the CNTs can be used for bright source of radiation generated through coherent nonlinear optical mixing [15,16]. Enhancement of efficiency in quantum-coherence based mechanisms of light generation is well-known [3,15,16]. Another reason of enhancement is the possibility to bypass the selection rule limitations: in a majority of cases, the systems with forbidden single-photon transitions are free of these limitations for multiphoton processes. In the case of carbon nanotubes, we can create an efficient tunable source of coherent radiation, which can potentially cover the frequency regions far from the band-gap energies of SWNTs. In particular, for the case of the double-Lambda scheme ((B) in Fig.1), we can suggest an experiment on the generation of single photons of IR radiation using coherent preparation of SWNTs by beams $\Omega_1$ and $\Omega_2$, and probing the system by the beam $\Omega_3$. As a result, we should generate IR radiation $\Omega_4$. This experiment will open up new opportunities in investigations of dark exciton states, provide an alternative way of determining bright-dark splitting energies, and help to reveal other features of dark exciton states in SWNTs. To a larger extent, the proposed experiment may allow us to solve the important problem of carbon nanotube optoelectronics- the creation of efficient carbon nanotube-based source of coherent radiation.

Other proposed quantum-optical schemes (see Fig.1) also may lead to new promising applications of SWNTs in optics and optoelectronics. In particular, the N-scheme is good for quantum computing, slow-fast light, quantum control, and achievement of a modified positive and negative index of refraction [17,20]. Use of the N scheme allow us to obtain entanglement between "bright" and "dark" states and monitor, maintain, or change the degree of coupling [25] between "dark" and "bright" states by external light pulses. It is important to point out that in carbon nanotubes, due to high binding energies of excitons (usually at least several times higher than the energies in nanosystems on the basis of "classical" semiconductors), all these effects have a chance to be realized at elevated temperatures.

An individual SWNTs can be electrically contacted and implemented into different electron transport devices [21,22] ( see also Fig.2). SWNTs look like a promising system (after quantum dot systems [26]) where quantum coherence may be controlled and modified not only via optical beams, but also via applied/measured voltages and currents. SWNTs can be a good "interface" between electronics and "classical", "all-optical" quantum optics. In addition, we would not rule out the possibility of using colloidal suspensions of SWNTs in different liquids and may be even in some solid matrices: in the case of reasonable separation of nanotubes and in the case of reasonably small dispersion of chiralities and diameters, the effect described above may also be observed and used.

To summarize, we report on proposed experiments demonstrating the quantum coherence effects on CNTs at various configurations ranging from the single-photon level, to the development of methamaterials using CNTs. We have shown that CNTs are a new material for quantum optics. Many applications of the coherent effects known for atomic and molecular media can be realized in CNTs. Even more, because the ratio of the relaxation rates in the exciton system is favorable, these effects can be even stronger than the effects in the well-known coherent media. The attractive feature of carbon nanotubes is the possibility of combining quantum-optical and electrotransport methods for manipulations with quantum states.

We thank Dr.Steve K. Doorn and Dr. Han Htoon for fruitful and inspiring discussions, and gratefully acknowledge the support from the NSF grant EEC-0540832 (MIRTHEERC), the Office of Naval Research, the Robert A. Welch Foundation (Grant #A1261), and the LANL-NMT MOU program supported by UCDRD.

REFERENCES

1. S. Reich, C. Thomse, and J. Maultzsch, *Carbon Nanotubes* (Wiley-VCH, 2004).

2. S.E.Harris, "Electromagnetically induced transparency," Physics Today 50,36-42 (1997).

3. M. O. Scully and M. S. Zubairy, *Quantum Optics*,(Cambridge University Press, 1997).

4. M. J. O'Connell, S. M. Bachilo, C. B. Huffman, V. C. Moore, M.S. Strano, E. H. Haroz, K. L. Rialon, P. J. Boul, W. H. Noon, C.Kittrell, J. Ma, R. H. Hauge, R. B. Weisman, and R. E. Smalley, "Band Gap Fluorescence from Individual Single-Walled Carbon Nanotubes," Science 297, 593-596 (2002).

5. F. Wang, G. Dukovic, L. E. Brus, and T. F. Heinz, "The Optical Resonances in Carbon Nanotubes Arise from Excitons," Science 308,838-841 (2005).

6. H. Zhao, S. Mazumdar C.-X. Sheng, M. Tong, and Z. V. Vardeny, "Photophysics of excitons in quasi-one-dimensional organic semiconductors: Single-walled carbon nanotubes and $\pi$-conjugated polymers," Phys. Rev. B 73, 075403 (2006), and references therein.

7. H.Kishida, Y.Nagasawa, S.Imamura, and A.Nakamura, "Direct Observation of Dark Excitons in Micelle-Wrapped Single-Wall Carbon Nanotubes," Phys Rev Lett 100, 097401, (2008), and references therein.

8. A.Srivastava, H.Htoon, V.I.Klimov, and J.Kono, "Direct observation of dark excitons in individual carbon nanotubes: role of local environments," http://arxiv.org/abs/0804.0875v1, and references therein.

9. E. B. Barros, R.B. Capaz, A.Jorio, G.G. Samsonidze, A.G. Souza Filho, S. Ismail-Beigi, C. D. Spataru, S. G. Louie, G.Dresselhaus, and M.S. Dresselhaus, "Selection rules for one- and two-photon absorption by excitons in carbon nanotubes," Phys. Rev. B 73, 241406 (2006).

10. V. Perebeinos, J. Tersoff, and Ph. Avouris, "Radiative Lifetime of Excitons in Carbon Nanotubes," http://arxiv.org/abs/cond-mat/0506775, and references therein.

11. S.Matsumoto, H.Matsui, A.Maeda, T.Takenobu, Y.Iwasa, Y.Miyata, H.Kataura, Y.Maniwa, and H. Okamoto, "Optical Stark Effect of Exciton in Semiconducting Single-Walled Carbon Nanotubes," Jap.Journ of Appl.Phys. 45 (20), L513-L515 (2006).

12. M.Fleischhauer, A.Imamoglu, J.P.Marangos, "Electromagnetically induced transparency: Optics in coherent media," Rev.Mod.Phys. 77, 633-673 (2005).

13. O.Kocharovskaya, Ya.I.Khanin, "Coherent population trapping and the attendant effect of absorptionless propagation of ultrashort pulse trains in a 3-level medium,", ZhETPh 90, 1610-1618 (1986).

14. V.A.Sautenkov, Y.V.Rostovtsev, C.Y.Ye, G.R.Welch, O.Kocharovskaya, and M.O.Scully, "Electromagnetically induced transparency in rubidium vapor prepared by a comb of short optical pulses," Phys. Rev. A 71, 063804 (2005).

15. N.G.Kalugin, Y.Rostovtsev, "Efficient generation of short terahertz pulses via stimulated Raman adiabatic passage," Opt.Lett.31, 969-971 (2006), and references therein.

16. M. Jain, H.Xia, G. Y. Yin, A. J. Merriam, and S. E. Harris, "Efficient Nonlinear Frequency Conversion with Maximal Atomic Coherence," Phys.Rev.Lett.77,4326-4329 (1996).

17. C.Liu, Z.Dutton, C.H.Behroozi, L.V.Hau, "Observation of coherent optical information storage in an atomic medium using halted light pulses," Nature,409,490-493 (2001).

18. S.E.Harris,Y.Yamamoto, "Photon Switching by Quantum Interference," Phys. Rev. Lett.81,3611-3614 (1998).

19. S.E.Harris and L.V. Hau, "Nonlinear Optics at Low Light Levels," Phys. Rev. Lett. 82, 4611-4614 (1999).

20. P.W. Milonni, *Fast light, slow light and left-handed light* ( Institute of Physics, 2005).

21. G. Fedorov, A. Tselev, D. Jiménez, S.Latil, N.G. Kalugin, P.Barbara, D.Smirnov, and S. Roche, "Magnetically Induced Field Effect in Carbon Nanotube Devices," Nano Lett. *7* (4), 960–964 (2007).

22. J. U. Lee, P. P. Gipp, and C. M. Heller, "Carbon nanotube *p-n* junction diodes," Appl. Phys. Lett. 85, 145-148 (2004).

23. H. Htoon, M. J. O'Connell, S. K. Doorn, and V. I. Klimov, "Single Carbon Nanotubes Probed by Photoluminescence Excitation Spectroscopy: The Role of Phonon-Assisted Transitions," Phys.Rev. Lett. 94, 127403 (2005).

24. A. Maeda, S. Matsumoto, H. Kishida, T. Takenobu, Y. Iwasa, M. Shiraishi, M. Ata, and H. Okamoto, "Large Optical Nonlinearity of Semiconducting Single-Walled Carbon Nanotubes under Resonant Excitations," Phys. Rev. Lett. 94, 047404 (2005), and references therein.

25. M. V. Gurudev Dutt, L. Childress, J. Liang, E. Togan, J. Maze, F. Jelezko, A. S. Zibrov, P.R. Hemmer, and M.D. Lukin,"Quantum Register Based on Individual Electronic and Nuclear Spin Qubits in Diamond," Science 316, 1312-1316 (2007).

26. A.V. Akimov, A. Mukherjee, C. L. Yu, D. E. Chang, A. S. Zibrov, P. R. Hemmer, H. Park, and M. D. Lukin, "Generation of Single Optical Plasmons in Metallic Nanowires Coupled to Quantum Dots," Nature 450, 402-406 (2007).

FIGURE CAPTIONS

Fig. 1. Energy diagrams and some of the possible schemes of quantum coherence preparation in the exciton system of semiconductor single-walled nanotubes. An excited electron band in use is labeled by $E_{jj}$ ($j$=1,2,..). Here (A) is one of the possible Lambda-schemes, (B) is the double-Lambda scheme , (C) the N-scheme, and (D) is the Ladder-Lambda scheme. The positions of the states are not exact. Symbols on the diagrams indicate the following: $\Omega_{1,...,4}$- are the Rabi frequencies of optical beams involved in the preparation of coherence and/or representing new radiation generated in a process, $\Delta_{1,2}$ are the values of detunings from a free electron state or from a bright exciton state correspondingly. In the case (A) ( see the details of discussion in the text) the generated coherence may be monitored via modulation of absorption or via modulation of intensity of "bright" exciton photoluminescence.

Fig.2. Possible experimental implementation of an SWNT excitonic quantum-optical device (a light modulator, detector, or emitter). Here, 1 is a suspended SWNT, 2 is a trench-structurized substrate (for example a $SiO_2$-coated heavily doped Si which also works as a backgate for SWNT), 3- source/drain and potential contacts to SWNT; 4-4 and 5-5 – mirrors forming optical cavities for corresponding quantum coherence-preparing and/or pump/probe optical beams. These mirrors should be positioned, oriented, and aligned in accordance with the particular variant of selected coherence-preparation scheme, with the diameter and chirality of the SWNT, and in accordance with the momentum conservation law for participating photons. Logically, a SWNT should be positioned in the areas of maximal focusing of participating beams. Therefore, the mirrors may be located around a SWNT transistor or even directly integrated with such a transistor and positioned in a Si/$SiO_2$ substrate trench. Alternative variants of device construction could be either just a SWNT field-effect or bipolar transistor [21,22] interacting with free-propagating beams focused on the SWNT [23], or different variants of constructions using surface plasmons (metallic stripes or nanoparticles positioned in the vicinity of a SWNT ). In the case of using plasmonic structures, the design of the device may be more complicated because of the necessity to take dispersion-related effects into account.

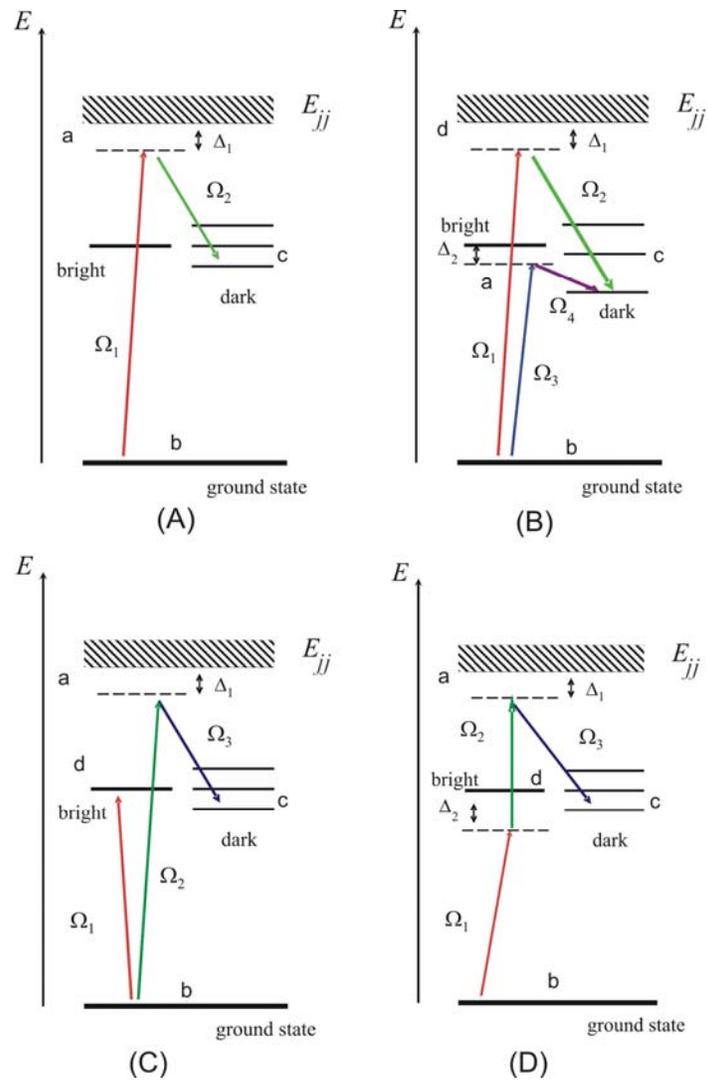

Figure 1.

Figure 2.